\documentclass[secnumarabic,amssymb, nobibnotes, aps, prd,showkeys]{revtex4-2}

\usepackage{hyperref}
\usepackage{graphicx}
\usepackage{caption}

\begin{document}

\title{Beauty Hadron Spectrum in a Screened Potential Model}%

\author{Raghavendra Kaushal}%
\email[Raghavendra Kaushal: ]{kaushalraghavendra23@gmail.com}
\affiliation{Department of Physics, Manipal Institute of Technology, Manipal Academy of Higher Education, Manipal, Karnataka, India, 576104}
\author{Bhaghyesh}%
\email[Bhaghyesh (Corresponding author): ]{bhaghyesh.mit@manipal.edu}
\affiliation{Department of Physics, Manipal Institute of Technology, Manipal Academy of Higher Education, Manipal, Karnataka, India, 576104}

\begin{abstract}
The mass spectrum of beauty hadrons ($b\overline{b}$ and $bbb$ baryons) and $bb$-diquarks are computed in a non-relativistic phenomenological potential model. The potential comprises of a short-range Coulomb potential, a screened confinement potential, and $O(1/m)$ corrections predicted from lattice and pNRQCD studies. Among the spin-dependent interactions, spin-spin interaction is considered non-perturbatively, whereas spin-orbit and tensor interactions are considered perturbatively. The Matrix-Numerov method is used to numerically solve the non-relativistic Schrodinger equation to evaluate the mass spectra. We interpret $\Upsilon(10753)$ as $D$-wave bottomonium state and $\Upsilon(10860)$ and $\Upsilon(11020)$ as $S$-wave bottomonium states. The mass spectrum of $bbb$ baryons are evaluated under the diquark-quark model. The excited masses are computed by considering various radial and orbital excitations of the diquark as well as the diquark-quark system.
\end{abstract}


\maketitle
\section{Introduction}\label{sec1}
The study of beauty hadrons (B-hadrons) is an interesting area of research as it allows one to explore the interplay between perturbative and non-perturbative quantum chromodynamics \cite{1,2}. It also plays a crucial role in determining the matrix elements of the CKM (Cabibbo-Kobayashi-Maskawa) matrix \cite{3,4}. The discovery of $\Upsilon(1S)$, a bottomonium state ($b\overline{b}$) in the year $1977$ marked the beginning of a new field called B-physics. $\Upsilon(1S)$ was observed by E288 Collaboration (Fermilab) while analyzing the spectrum of dimuons ($\mu^+ \mu^-$) produced by the collision of a protons with Pb and Cu targets at $400$ GeV \cite{5,6}. The orbitally excited bottomonium states $\chi_{bJ}(1P)$ were confirmed by analyzing the electromagnetic spectrum of $\Upsilon(2S)$ obtained at DESY \cite{7}. CLEO collaboration \cite{8} presented the first evidence for the existence of D-wave bottomonium state $\Upsilon(1D)$ by analyzing the decays, $\Upsilon(3S) \rightarrow \gamma \chi_{b}(2P_J), \chi_{b}(2P_J) \rightarrow \gamma \Upsilon(1D)$. In the last three decades, the bottomonium sector has evolved exceptionally due to the extraordinary efforts by various collaborations like CLEO collaboration \cite{8,9,10,11}, Belle collaboration \cite{12,13,14}, BaBar collaboration \cite{15,16,17,18}, ATLAS collaboration \cite{20}, CMS collaboration \cite{21}, and LHCb collaboration \cite{22} to name few. In 2007, the CDF collaboration \cite{22a} reported the first observation of beauty baryons produced at the Tevatron from the proton-antiproton collisions at $1.1$ fb$^{-1}$. They interpreted the discovered states as $\Sigma_b^{+}$ with mass $5807.8^{+2.0}_{-2.2}\pm1.7$ MeV/c$^2$ and $\Sigma_b^{-}$ with mass $5815.2 \pm 1.0 \pm 1.7$ MeV/c$^2$. Currently, we have several baryons that consist of a single bottom quark ($\Lambda_b^0, \Sigma_b^0, \Sigma_b^+, \Sigma_b^-, \Xi_b^0, \Xi_b^-, $ and $\Omega_b^-$), but baryons consisting of two or three bottom quarks have not been discovered yet \cite{23}. The authors of Ref. \cite{24} computed the fragmentation functions for triply charmed ($\Omega_{ccc}$) and triply bottom ($\Omega_{bbb}$) baryons by applying the perturbation theory in Feynman gauge by incorporating contributions from two Feynman diagrams. They obtained the fragmentation probability for $\Omega_{ccc}$ and $\Omega_{bbb}$ of the order of $10^{-9}$, which they believe can be observed at LHC because of its high collision rates. However, the authors of Ref. \cite{25} argue that this computed fragmentation function is not accurately described as there will be a minimum of seven Feynman diagrams that would contribute to the fragmentation function. Nevertheless, their calculations show that at an integrated luminosity of $10$ $fb^{-1}$ at LHC, there could be $10^4$ to $10^5$ events of triply heavy baryons and concluded that it is quite favorable for triply heavy baryons to be discovered at LHC. Investigating bottom-quark physics is essential as it allows us to examine the flavor dynamics of the Standard Model and may uncover the reasons behind the existence of three generations of particles in nature. The study of B-hadrons are regarded as an important pathway for grasping the fundamental principles of particle physics \cite{75}. Due to the higher mass of bottom quarks compared to charm or other light quarks, the behavior of bottom quarks in the quark-gluon plasma changes from losing energy to diffusion. Examining beauty hadrons can offer crucial information about the shift from high-energy parton movement to low-energy diffusion, which plays a key role in comprehending the dynamics of the quark-gluon plasma \cite{76}. 

Various theoretical approaches such as effective field theories \cite{26,27,28,29}, lattice QCD \cite{30,31,32}, and phenomenological potential models \cite{33,34,35,36} are used to study the spectra of beauty hadrons. The theoretical study of triply bottom baryons is anticipated to provide essential insights for future experimental endeavors. Various theoretical approaches employed to investigate the spectra of triply bottom baryons include the constituent quark model \cite{37,38,79,80}, QCD sum rules \cite{39,40,81}, diquark-quark model \cite{33,41}, lattice QCD \cite{32,42}, and the hypercentral constituent quark model \cite{43,82}. The existing literature suggests that there has been insufficient investigation into the excited states of triply bottom baryons. The quark models effectively describe various properties of baryons, such as mass spectra, magnetic moments, form factors, strong decays, and so on. Nonetheless, these models predict a greater number of states than what has been observed in experiments (for lighter baryons). This discrepancy is known as the missing resonance problem \cite{89,90}. An alternative solution to this problem is to use models that utilize less number of effective degrees of freedom compared to the traditional three-quark model for baryons. This has led to the development of the quark-diquark model for baryons \cite{45}. In this work, we employ the diquark-quark approach to investigate the mass spectrum of triply bottom baryons. Diquark, as the name suggests, is a (hypothetical) bound state of two quarks \cite{44,45,46}. Diquarks are colored and hence cannot be found as an isolated particle. The significance of modelling baryons as a bound state of diquark and quark is that the color interaction between diquark and quark is similar to that of an antiquark and quark in mesons \cite{47} In this work, we compute the mass spectra of bottomonium, $bb$-diquarks, and triply bottom baryons utilizing a non-relativistic phenomenological potential model with one-gluon Coulomb-like short-range potential and the screened confinement potential. In the literature, we observed that most of the phenomenological potential models use linear confinement potential. It is because the linear confinement potential can be obtained from Yang-Mills gauge theories and also certain lattice results support this \cite{48,49}. There are various other confinement potentials used in the literature (for detailed review refer \cite{50}). We will be utilizing the screened confinement potential, as it has been found that with the increasing separation between the quarks, dynamic quark-antiquark pairs may emerge from the vacuum, leading to vacuum polarization. As a result of this, the interquark potential will be subject to screening \cite{48,51}. Furthermore, we incorporate the $O(1/m)$ corrections derived from LQCD \cite{52} and pNRQCD \cite{26}, which are simulated using the quenched approximation. These corrections are attributed to the nonzero masses of the heavy quarks. Our work integrates quenched $O(1/m)$ corrections with a screened confinement potential, establishing a link between the theoretical frameworks influenced by quenched QCD and practical representations that take into account the unquenched effects.

The paper is structured as follows. In section \ref{S2}, we have given the theoretical framework where we have discussed the potential model and the method employed to numerically solve the Schrodinger wave equation to evaluate the mass spectra of beauty hadrons ($b\overline{b}$ and $bbb$ baryon). In section \ref{S3}, we have presented the mass spectra of beauty hadrons and have compared with other theoretical approaches and available experimental results. In section \ref{S4}, we have provided our conclusions.

\section{Theoretical Framework: \label{S2}}
In systems consisting of heavy quarks, the kinetic energy of the individual quarks is negligible compared to their rest mass energy. Hence it is favorable to model these systems using a non-relativistic approach. To evaluate the mass spectra, we make use of nonrelativistic radial Schrodinger equation given by, 
\begin{eqnarray}
	\left[\frac{1}{2\mu}\left(-\frac{d^2}{d r^2}+\frac{l(l+1)}{r^2}\right)+V(r)\right]u(r)=Eu(r).
	\label{E1}	
\end{eqnarray}
Here $\mu$ corresponds to the reduced mass of the system, $\mu=(m_1 m_2)/(m_1+m_2)$, $l$ to the relative orbital angular momentum of the system, and $r$ to the interquark seperation respectively. Equation (\ref{E1}) will be solved numerically. In the literature, we can find various numerical methods to solve the Schrodinger equation. Prominent among them are Discrete Variable Representation (DVR) method  \cite{53}, Nikiforov-Uvarov (NU) method \cite{54}, Runge-Kutta Method \cite{55}, Matrix Numerov Method \cite{56}, Fourier grid Hamiltonian method \cite{57}, Variational method \cite{58}, Phase Integral method \cite{59}, and Asymptotic Iteration method \cite{60}. In this work, we use the Matrix Numerov method because of its efficiency in simultaneously computing multiple eigenstates \cite{56}. The method is briefly outlined as follows:

Equation (\ref{E1}) is rewritten as,
\begin{eqnarray}
	\frac{d^2u(r)}{d r^2}=-\left(2\mu E-V(r)-\frac{l(l+1)}{r^2}\right)u(r)
	=g(r)u(r).
	\label{E2}
\end{eqnarray}
with $g(r)=-\left(2\mu E-V(r)-\frac{l(l+1)}{r^2}\right)$.
The space will be discretized into $N$ equal parts that are seperated by a distance $d$. The second derivative of $u(r)$ in equation (\ref{E2}) will be approximated using Taylor series expansion of $u(r)$.
\begin{eqnarray}
	u(r\pm d)=u(r)\pm du^{(1)}(r)+\frac{1}{2!} d^{2}u^{(2)}(r) 
	\pm \frac{1}{3!}d^3u^{(3)}(r)
	+\frac{1}{4!}d^4u^{(4)}(r)+\dots
	\label{E3}
\end{eqnarray}
with $u^{(n)}(r)=\frac{d^n u(r)}{dr^n}$.
From equation (\ref{E3}), the second derivative of $u(r)$ can be written as,
\begin{eqnarray}
	u^{(2)}(r)=\frac{u(r+d)+u(r-d)-2u(r)}{d^2} 
	-\frac{1}{12}d^2u^{(4)}(r)+O(d^4)
	\label{E4}
\end{eqnarray}
Substituting equation (\ref{E4}) in equation (\ref{E2}) and rearranging, we get,
\begin{eqnarray}
	u^{(4)}(r)=\frac{g(r+d)u(r+d)+g(r-d)u(r-d)-2g(r)u(r)}{d^2} 
	+O(d^2)
	\label{E5}
\end{eqnarray}
Substituting equation (\ref{E5}) in equation (\ref{E4}) and rearranging, we get,
\begin{eqnarray}
	g_ju_j=\frac{u_{j+1}+u_{j-1}-2u_j}{d^2}-\frac{1}{12}\left(g_{j+1}u_{j+1}+g_{j-1}u_{j-1}-2g_ju_j\right)
	\label{E6}
\end{eqnarray}
Here $g_{j-1}=g(r-d)$, $g_j=g(r)$, $g_{j+1}=g(r+d)$, $u_{j-1}=u(r-d)$, $u_j=u(r)$, $u_{j+1}=u(r+d)$.
By comparing equations (\ref{E6}) and (\ref{E2}), the Schrodinger equation in matrix form can be written as \cite{56,bb},

\begin{eqnarray}
	\left[-\frac{1}{2\mu}(AB^{-1})+\frac{l(l+1)}{2\mu r_j^2}+V(r_j)\right]u_j=Eu_j.
	\label{E7}
\end{eqnarray}
with $u_j=u(r_j)$ is the wavefunction at $r_j$, $r_j=r_{min}+jd$, $d=(r_{max}-r_{min})/N$, $r_{max}$ and $r_{min}$ are the upper and lower bounds of the space under consideration, $d$ is the distance between the two discretized points, $A=(I_{-1}-2I_{0}+I_{1})/d^2$, $B=(I_{-1}+10I_0+I_1)/12$, $I_{-1}$, $I_0$, and $I_1$ are lower, diagonal unit matrix, and upper shift matrix respectively i.e.

\begin{eqnarray}
	I_{-1}=
	\left(
	\begin{array}{ccc}
		0 & 0 & \cdots \\
		1 & 0 & \cdots \\
		\vdots & 1 & \ddots
	\end{array}
	\right)
	,  I_{0}=
	\left(
	\begin{array}{cccc}
		1 & 0 & 0 & \cdots \\
		0 & 1 & 0 & \cdots \\
		\vdots & 0 & 0 & \ddots
	\end{array}
	\right)
	,   I_{1}=
	\left(
	\begin{array}{cccc}
		0 & 1 & 0 & \cdots \\
		0 & 0 & 1 & \cdots \\
		\vdots & 0 & 0 & \ddots
	\end{array}
	\right)
\end{eqnarray}
Equation (\ref{E7}) is solved using Wolfram Mathematica software. In this work, we take $N=500$ and $r_{max}=5$ fm.

Typically, the potentials in phenomenological potential models consists of both spin-independent and spin-dependent parts. Spin-dependent interaction comprises of spin-spin, spin-orbit, and tensor interactions. In the literature, we can observe that in most of the models, the spin-dependent interactions have been considered perturbatively. Nevertheless, it has been suggested in Refs \cite{61,62} that the spin-spin interaction plays an important role in the binding energy, making it more favorable to treat it in a non-perturbative manner. Therefore, we will be considering the spin-spin interaction non-perturbatively whereas spin-orbit and tensor interactions will be considered perturbatively. Spin-independent interaction usually comprises of one-gluon short-range Coulomb-like potential and the confinement potential. The potential that we will be employing to solve Schrodinger equation (\ref{E7}) includes one-gluon short-range Coulomb-like potential, screened confinement potential, $O(1/m)$ corrections, and the spin-spin interaction given by,
\begin{eqnarray}
	\label{E9}
	V(r)=\frac{\kappa\alpha_{s}(Q^2)}{r} +\frac{\lambda(1-e^{-\nu r})}{\nu}+\left(\frac{1}{m_1} +\frac{1}{m_2}\right)\left(\frac{-9\alpha^{2}}{8r^2}+C\ln(ar)\right)+V_{SS}(r).
\end{eqnarray}

Here $\kappa$ corresponds to color factor, $\lambda$ is the string strength, $\nu$ is the screening factor, $\alpha_{s} (Q^2)$ is the QCD coupling constant at the renormalization scale $Q$, $\alpha$ is the effective coupling constant given by $\alpha=\kappa \alpha_{s}(Q^2)$, $a$ and $C$ are phenomenological constants, $r$ is the interquark distance, $m_1$ and $m_2$ are the masses of constituent particles, and $V_{SS}(r)$ is the spin-spin interaction respectively. $\alpha_{s}(Q^2)$ is obtained by using the formula \cite{cc},
\begin{eqnarray}
	\alpha_s(Q^2)=\frac{4\pi}{\left(11-\frac{2}{3}n_f\right)ln(\frac{Q^2}{\Lambda_{QCD}^2})}.
	\label{E10}
\end{eqnarray}
where $\Lambda_{QCD}$ is the QCD scale whose value is around $0.1$ to $0.4$ GeV \cite{23}. In this work, we have used $n_f=4$ and $\Lambda_{QCD}=0.130 $ GeV. In the case of bottomonium, the renormalization scale $Q$ will be equal to the mass of bottom quark ($m_b$). We will be using this same renormalization scale ($Q=m_b$) for $bb$ diquark and triply bottom baryons.

The potential parameters are determined by considering the masses of $8$ experimentally well-established bottomonium states ($\eta_b(1S)$, $\Upsilon(1S)$, $\eta_b(2S)$, $\Upsilon(2S)$, $\chi_{b0}(1P)$, $\chi_{b1}(1P)$, $h_b(1P)$, and $\chi_{b2}(1P)$) and by minimizing the chi-square, $ \chi^2=\sum_{i=1}^n \left((M_i^{exp}-M_i^{cal})/\Delta M_i^{exp}\right)^2$, where $M_i^{exp}$ and $M_i^{cal}$ are the experimental and calculated masses of the $i^{th}$ bottomonium state respectively and $\Delta M_i^{exp}$ is the experimental uncertainty in the mass of $i^{th}$ bottomonium state. We have taken a constant value for the experimental uncertainty, $\Delta M_i^{exp}=5$ MeV as suggeted in Ref \cite{aa}. This decision has been made due to the small and inconsistent experimental uncertainties in the masses of the states used for fitting. If the experimental uncertainty is excessively low, it will result in an overly precise mass, which can be unfavorable for fitting purposes \cite{aa}. The fitted potential parameters are shown in Table \ref{T1}.

\begin{table}
	\caption{\label{T1} Model Parameters.}
	\begin{ruledtabular}
		\begin{tabular}{@{}cccccc}
			$m_{b}$ [GeV]&$\lambda$ [GeV$^2$]&$\nu$ [GeV]&$C$ [GeV]&$a$ [GeV]&$\sigma$ [GeV]\\
			\hline
			$4.680$&$0.241$&$0.078$&$0.100$&$0.430$&$3.920$\\
		\end{tabular}
	\end{ruledtabular}
\end{table}

\subsection{Bottomonium:}
The color factor in equation (\ref{E9}) for bottomonium, which is a bound state of bottom and anti-bottom quarks will be $\kappa=-\frac{4}{3}$. Hence, the potential (\ref{E9}) for bottomonium takes the form,
\begin{eqnarray}
	V(r)=-\frac{4}{3} \frac{\alpha_{s}(m_b^2)}{r}+\frac{\lambda(1-e^{-\nu r})}{\nu}+\left(\frac{1}{m_b} +\frac{1}{m_b}\right)\left(\frac{-9\alpha^{2}}{8r^2}+C\ln(ar)\right)\nonumber\\
	+\frac{32\pi\alpha_s(m_b^2)}{9m_b^2}\left(\frac{\sigma}{\sqrt{\pi}}\right)^3exp(-\sigma^2 r^2) \mathbf{S_1}\cdot\mathbf{S_2}.
	\label{E11}
\end{eqnarray}

In equation (\ref{E11}), the fourth term corresponds to smeared spin-spin interaction. Additionally, the other spin-dependent interactions such as spin-orbit and tensor interactions are given respectively as  \cite{87},

\begin{eqnarray}
	V_{LS}(r)=\frac{1}{2m_b^2 r}\left(3 \frac{d V_V(r)}{d r}-\frac{d V_S(r)}{d r}\right) \mathbf{L}\cdot \mathbf{S}.
	\label{E12}
\end{eqnarray}
\begin{eqnarray}
	V_T(r)=\frac{1}{12m_b^2}\left(\frac{1}{r}\frac{d V_V(r)}{d r}-\frac{d^2 V_V(r)}{d r^2}\right) S_{12}.
	\label{E13}
\end{eqnarray}
where,
\begin{eqnarray}
	S_{12}=12\left((\mathbf{S}_1\cdot\hat{r})(\mathbf{S}_2\cdot\hat{r})-\frac{1}{3}\mathbf{S}_1\cdot\mathbf{S}_2\right).
	\label{E14}
\end{eqnarray} 
Here $V_{V}(r)$ is a lorentz vector that corresponds to short-range interquark potential and $V_S(r)$ is a lorentz scalar that corresponds to the confinement potential. In this work, we take,
\begin{eqnarray}
	V_V(r)=\frac{\kappa \alpha_{s}}{r}, \quad
	V_S(r)=\frac{\lambda (1-e^{-\nu r})}{r}
\end{eqnarray}

The masses of bottomonium states will be computed as,
\begin{eqnarray}
	M(b\bar{b})=E+\langle V_{LS}(r)\rangle+\langle V_T(r)\rangle+2m_b.
	\label{E16}
\end{eqnarray}
where $E$ corresponds to the eigenvalue (binding energy) obtained by solving the Schrodinger equation (\ref{E7}), $\langle V_{LS}(r)\rangle$ and $\langle V_T(r)\rangle$ corresponds to perturbatively determined spin-orbit and tensor interaction energies respectively, and $m_b$ corresponds to mass of bottom (antibottom) quark. The expectation values in equation (\ref{E16}) are evaluated using the wavefunctions obtained by solving the Schrodinger equation (\ref{E7}) for the potential given by equation (\ref{E11}).

\subsection{$bb$ diquarks:}
In the fundamental color representation, quarks are represented by a color triplet $\mathbf{3}$ state \cite{64}. The two quarks combine each other as $\mathbf{3} \otimes \mathbf{3} =\mathbf{\overline{3}} \oplus \mathbf{6}$. Here $\mathbf{\overline{3}}$ represents a color antitriplet state and $\mathbf{6}$ represents a color sextet state. The diquarks can either be in a color antitriplet state or in a color sextet state. Lets assume that the diquark is in color sextet state. The diquark combines with a quark as $\mathbf{6} \otimes \mathbf{3} = \mathbf{10} \oplus \mathbf{8}$. Here $\mathbf{10}$ represents a color decuplet state and $\mathbf{8}$ represents a color octet state. Since baryons are physically observable, it must be a color singlet ($\mathbf{1}$) state. Hence, diquarks cannot be in a color sextet state as the combination with quark does not give a color singlet state. Now, lets assume that the diquark is in color antitriplet state. The diquark combines with a quark as $\mathbf{\overline{3}} \otimes \mathbf{3} = \mathbf{1} \oplus \mathbf{8}$. This combination does give a color singlet state. Hence, the diquark will be in the color antitriplet state. In the fundamental color representation, antiquark is represented by an antitriplet state \cite{64}. Therefore, this combination of antitriplet diquark with a quark in a baryon is identical to the combination of antitriplet antiquark with a quark in a meson. Hence, the color interaction between diquark and quark in baryon is identical to the color interaction between antiquark and quark in meson \cite{47} which is well studied and well understood. 

As $bb$ diquarks are made up of identical $b$ quarks, according to Pauli Exclusion principle, the total wavefunction of the diquark system must be antisymmetric \cite{45}. Parity of the diquark, given by $(-1)^{l_d}$ describes the nature of the spatial wavefunction, where $l_d$ is the relative orbital angular momentum of the diquark. Since the flavor of quarks under consideration are same, the flavor wavefunction is symmetric. For color antitriplet diquark state, the color wavefunction is antisymmetric \cite{47}. This suggests that for $bb$ diquark, the color-flavor wavefunction is antisymmetric. Therefore, inorder to have total wavefunction antisymmetric, the spin-spatial wavefunction must be symmetric.

For the diquark in the antitriplet color state, the color factor between the interaction of two quarks will be $\kappa=-\frac{2}{3}$ which is half the color factor between the interaction of a quark and an antiquark. It has been suggested in Refs \cite{61,65} that this factor of $\frac{1}{2}$ arises because of the color wavefunction and that it has to be considered for the entire static potential. Hence, we have the static potential for diquark as,
\begin{eqnarray}
	V_{qq}^{(0)}(r)=\frac{V_{q\bar{q}}^{(0)}(r)}{2}=-\frac{2}{3}\frac{\alpha_{s}(m_b^2)}{r}+\frac{\lambda (1-e^{-\nu r})}{2 \nu}
\end{eqnarray}
Including the $O(1/m)$ corrections and the spin-spin interactions, we get the diquark potential as,
\begin{eqnarray}
	V_{qq}(r)=-\frac{2}{3} \frac{\alpha_{s}(m_b^2)}{r}+\frac{\lambda(1-e^{-\nu r})}{2 \nu}+\left(\frac{1}{m_b} +\frac{1}{m_b}\right)\left(\frac{-9\alpha^{2}}{8r^2}+C\ln(ar)\right)\nonumber\\
	+\frac{16\pi\alpha_s(m_b^2)}{9m_b^2}\left(\frac{\sigma}{\sqrt{\pi}}\right)^3exp(-\sigma^2 r^2) \mathbf{S_1}\cdot\mathbf{S_2}.
\end{eqnarray}
The contributions from spin-orbit and tensor interactions will be computed perturbatively, using the formula given by equations (\ref{E12}) and (\ref{E13}).

\subsection{Triply bottom baryons:}
\begin{figure}
	\centering
	\includegraphics[width=0.5\linewidth]{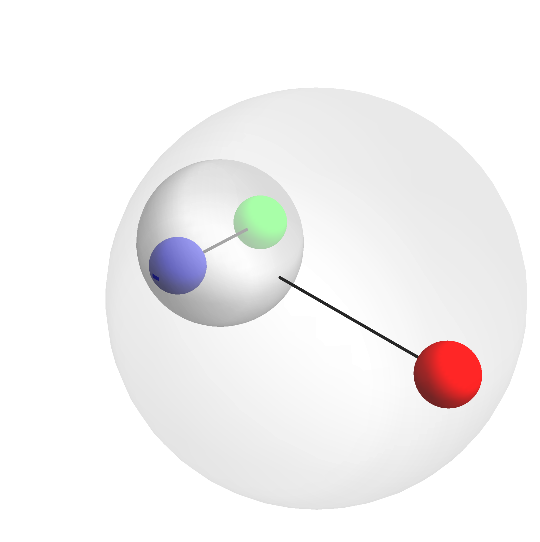} 
	\caption{Pictorial representation of baryon in diquark-quark model.}
	\label{F3}
\end{figure}

The mass spectrum of the triply bottom baryons are evaluated within the quark-diquark approach. The baryon is treated as a bound state of a $bb$-diquark and a $b$-quark, pictorially represented in Figure \ref{F3}. We consider the same model parameters as given in Table \ref{T1} for the evaluation of mass spectrum of triply bottom baryons. In this work, the baryon states will be represented as `$n_dl_dnL$' where $n_d$ and $l_d$ represents the radial and orbital quantum numbers  of the diquark system respectively and $n$ and $L$ represents the radial and orbital quantum numbers of diquark-quark system respectively. The total wavefunction of baryon is a combination of spin, flavor, spatial, and color wavefunctions given by,
\begin{eqnarray}
	\psi_{baryon}=\psi_{spin}\psi_{flavor}\psi_{spatial}\psi_{color}
\end{eqnarray}
The total wavefunction must be antisymmetric in accordance with Pauli Exclusion principle \cite{33}. For $bbb$ baryon, the flavor wavefunction will be symmetric, color wavefunction will be antisymmetric, requiring the spin-spatial wavefunction to be symmetric. The spatial wavefunction depends on the parity of the baryon given by $(-1)^{l_d+L}$ where $l_d$ is the relative orbital angular momentum of diquark and $L$ is the relative orbital angular momentum of diquark-quark system respectively.

The spin-dependent interactions for baryons under the diquark-quark approach are similar to spin-dependent interactions in heavy open-flavored mesons. The spin-dependent interactions are given by \cite{66,67,68},

\begin{eqnarray}
	V_{SS}(r)=\frac{32\pi\alpha_s}{9m_dm_q}\left(\frac{\sigma}{\sqrt{\pi}}\right)^3exp(-\sigma^2 r^2) \mathbf{J_d}\cdot\mathbf{S_q}.
	\label{E22}
\end{eqnarray}
\begin{eqnarray}
	V_{LS}(r)=\frac{4}{3}\alpha_s\frac{1}{m_d m_q r^3 }\mathbf{L}\cdot\mathbf{S}+\left(\frac{1}{2m_d^2}\mathbf{L}\cdot\mathbf{J_d}+\frac{1}{2m_q^2}\mathbf{L}\cdot\mathbf{S_q}\right)\frac{1}{r}\frac{d}{d r}\left(V_V(r)-V_S(r)\right).
	\label{E23}
\end{eqnarray}
\begin{eqnarray}
	V_T(r)=\frac{-4\alpha_s\left[(\mathbf{L}\cdot\mathbf{J_d})(\mathbf{L}\cdot\mathbf{S_q})+(\mathbf{L}\cdot\mathbf{S_q})(\mathbf{L}\cdot\mathbf{J_d})-\frac{2}{3}L(L+1)(\mathbf{J_d}\cdot\mathbf{S_q})\right]}{m_d m_q r^3 (2L-1)(2L+3)}. 
	\label{E24}
\end{eqnarray}
Here $m_q$ and $m_d$ corresponds to the masses of constituent quark and diquark respectively, $\mathbf{J_d}$ corresponds to total angular momentum of diquark given by, $\mathbf{J_d}=\mathbf{l_d}+\mathbf{S_d}$, $\mathbf{S_d}$ and $\mathbf{l_d}$ corresponds to total spin and relative orbital angular momentum of diquark respectively, $\mathbf{L}$ corresponds to relative orbital angular momentum of diquark-quark system, and $\mathbf{S_q}$ corresponds to spin of the quark interacting with diquark. In diquark-quark model, $\mathbf{J_d}$ will be considered as the spin of diquark \cite{34}. $\mathbf{S}$ corresponds to the total spin of baryon given by, $\mathbf{S}=\mathbf{S_q}+\mathbf{J_d}$. Application of Pauli Exclusion principle for diquarks results in $S$-wave diquark with $\mathbf{J_d}=1$, $P$-wave diquark with $\mathbf{J_d}=1$, and $D$-wave diquark with $\mathbf{J_d}=1,2,3$. The mass difference across different $\mathbf{J_d}$ values for $D$-wave diquarks were found to be negligible (refer Table \ref{T7}). Hence, for the computation of baryon masses, we consider $S$, $P$, and $D$-wave diquark with $\mathbf{J_d}=1$ (if we were to consider $D$-wave diquark with $\mathbf{J_d}=2,3$, there could be mixing between states). The spin-orbit and tensor interactions will be evaluated in $|J;S\rangle$ basis. The basis transformation relation between $|J;j_q\rangle$ and $|J;S\rangle$ basis is given by \cite{34},
\begin{eqnarray}
	|J;j_q\rangle=\sum_{S}(-1)^{J+L+J_d+S_q}\sqrt{(2S+1)(2j_q+1)}
	\left\{	\begin{array}{ccc}
		J_d & S_q & S \\
		L & J & j_q
	\end{array} \right\}|J;S\rangle.
	\label{E25}
\end{eqnarray}
In $|J;j_q\rangle$ basis, the coupling takes place as follows. First, the relative orbital angular momentum of diquark-quark system $\mathbf{L}$ couples with the spin of the quark $\mathbf{S_q}$ to form $\mathbf{j_q}$ i.e., $\mathbf{L}+\mathbf{S_q}=\mathbf{j_q}$. Then, $\mathbf{j_q}$ couples with the total spin of diquark $\mathbf{J_d}$ to form the total angular momentum $\mathbf{J}$ of baryon i.e.,  $\mathbf{j_q}+\mathbf{J_d}=\mathbf{J}$. In $|J;S\rangle$ basis, coupling takes place as follows. First, the total spin of diquark $\mathbf{J_d}$ couples with the spin of quark $\mathbf{S_q}$ to form the total spin of baryon $\mathbf{S}$ i.e., $\mathbf{J_d}+\mathbf{S_q}=\mathbf{S}$. Then, the total spin of baryon couples with the relative orbital angular momentum of diquark-quark system $\mathbf{L}$ to form the total angular momentum $\mathbf{J}$ of baryon i.e., $\mathbf{S}+\mathbf{L}=\mathbf{J}$. In equation (\ref{E25}), $|J;j_q\rangle$ basis is represented in terms of $|J;S\rangle$ basis. However, we will be using $|J;S\rangle$ basis represented in terms of $|J;j_q\rangle$ basis. We have given the basis states that are used in the evaluation of spin-orbit and tensor interactions in Table \ref{T2}. 

\begin{table}
	\caption{\label{T2}$|J;S\rangle$ wavefunctions in terms of $|J;j_q\rangle$.}
	\begin{ruledtabular}
		\begin{tabular}{@{}ccccccc}
			States&$L$&$S$&$J$&$j_q$&$|J;S\rangle$&$|J;j_q\rangle$\\
			\hline
			\vspace{6pt}
			$1S1p,2S1p,1S2p,2S2p$&$1$&$\frac{1}{2}$&$\frac{1}{2}$&$\frac{1}{2},\frac{3}{2}$&$|\frac{1}{2};\frac{1}{2}\rangle$&$\frac{-1}{3}|\frac{1}{2};\frac{1}{2}\rangle+\frac{2\sqrt{2}}{3}|\frac{1}{2};\frac{3}{2}\rangle$\\
			\vspace{6pt}
			&$1$&$\frac{1}{2}$&$\frac{3}{2}$&$\frac{1}{2},\frac{3}{2}$&$|\frac{3}{2};\frac{1}{2}\rangle$&$\frac{-2}{3}|\frac{3}{2};\frac{1}{2}\rangle+\frac{\sqrt{5}}{3}|\frac{3}{2};\frac{3}{2}\rangle$\\
			\vspace{6pt}
			$1S1d,2S1d,1S2d,2S2d$&$2$&$\frac{3}{2}$&$\frac{3}{2}$&$\frac{3}{2},\frac{5}{2}$&$|\frac{3}{2};\frac{3}{2}\rangle$&$\frac{2}{\sqrt{5}}|\frac{3}{2};\frac{3}{2}\rangle+\frac{1}{\sqrt{5}}|\frac{3}{2};\frac{5}{2}\rangle$\\
			\vspace{6pt}
			&$2$&$\frac{3}{2}$&$\frac{5}{2}$&$\frac{3}{2},\frac{5}{2}$&$|\frac{5}{2};\frac{3}{2}\rangle$&$\sqrt{\frac{7}{15}}|\frac{5}{2};\frac{3}{2}\rangle+2\sqrt{\frac{2}{15}}|\frac{5}{2};\frac{5}{2}\rangle$\\
			\vspace{6pt}
			$1P1p,2P1p,1P2p,2P2p$&$1$&$\frac{3}{2}$&$\frac{1}{2}$&$\frac{1}{2},\frac{3}{2}$&$|\frac{1}{2};\frac{3}{2}\rangle$&$\frac{2\sqrt{2}}{3}|\frac{1}{2};\frac{1}{2}\rangle+\frac{1}{3}|\frac{1}{2};\frac{3}{2}\rangle$\\
			\vspace{6pt}
			&$1$&$\frac{3}{2}$&$\frac{3}{2}$&$\frac{1}{2},\frac{3}{2}$&$|\frac{3}{2};\frac{3}{2}\rangle$&$\frac{\sqrt{5}}{3}|\frac{3}{2};\frac{1}{2}\rangle+\frac{2}{3}|\frac{3}{2};\frac{3}{2}\rangle$\\
			\vspace{6pt}
			$1P1d,2P1d,1P2d,2P2d$&$2$&$\frac{1}{2}$&$\frac{3}{2}$&$\frac{3}{2},\frac{5}{2}$&$|\frac{3}{2};\frac{1}{2}\rangle$&$\frac{-1}{\sqrt{5}}|\frac{3}{2};\frac{3}{2}\rangle+\frac{2}{\sqrt{5}}|\frac{3}{2};\frac{5}{2}\rangle$\\
			\vspace{6pt}
			&$2$&$\frac{1}{2}$&$\frac{5}{2}$&$\frac{3}{2},\frac{5}{2}$&$|\frac{5}{2};\frac{1}{2}\rangle$&$-2\sqrt{\frac{2}{15}}|\frac{5}{2};\frac{3}{2}\rangle+\sqrt{\frac{7}{15}}|\frac{5}{2};\frac{5}{2}\rangle$\\
			\vspace{6pt}
			$1D1p,2D1p,1D2p,2D2p$&$1$&$\frac{1}{2}$&$\frac{1}{2}$&$\frac{1}{2}, \frac{3}{2}$&$|\frac{1}{2};\frac{1}{2}\rangle$&$\frac{-1}{3}|\frac{1}{2};\frac{1}{2}\rangle+\frac{2\sqrt{2}}{3}|\frac{1}{2};\frac{3}{2}\rangle$\\
			\vspace{6pt}
			&$1$&$\frac{1}{2}$&$\frac{3}{2}$&$\frac{1}{2},\frac{3}{2}$&$|\frac{3}{2};\frac{1}{2}\rangle$&$\frac{-2}{3}|\frac{3}{2};\frac{1}{2}\rangle+\frac{\sqrt{5}}{3}|\frac{3}{2};\frac{3}{2}\rangle$\\
			\vspace{6pt}
			$1D1d,2D1d,1D2d,2D2d$&$2$&$\frac{3}{2}$&$\frac{3}{2}$&$\frac{3}{2},\frac{5}{2}$&$|\frac{3}{2};\frac{3}{2}\rangle$&$\frac{2}{\sqrt{5}}|\frac{3}{2};\frac{3}{2}\rangle+\frac{1}{\sqrt{5}}|\frac{3}{2};\frac{5}{2}\rangle$\\
			\vspace{6pt}
			&$2$&$\frac{3}{2}$&$\frac{5}{2}$&$\frac{3}{2},\frac{5}{2}$&$|\frac{5}{2};\frac{3}{2}\rangle$&$\sqrt{\frac{7}{15}}|\frac{5}{2};\frac{3}{2}\rangle+2\sqrt{\frac{2}{15}}|\frac{5}{2};\frac{5}{2}\rangle$\\
		\end{tabular}
	\end{ruledtabular}
\end{table}

\section{Results and Discussions: \label{S3}}

\subsection{Bottomonium:}
\begin{figure}
	\centering
	\includegraphics[width=1\linewidth]{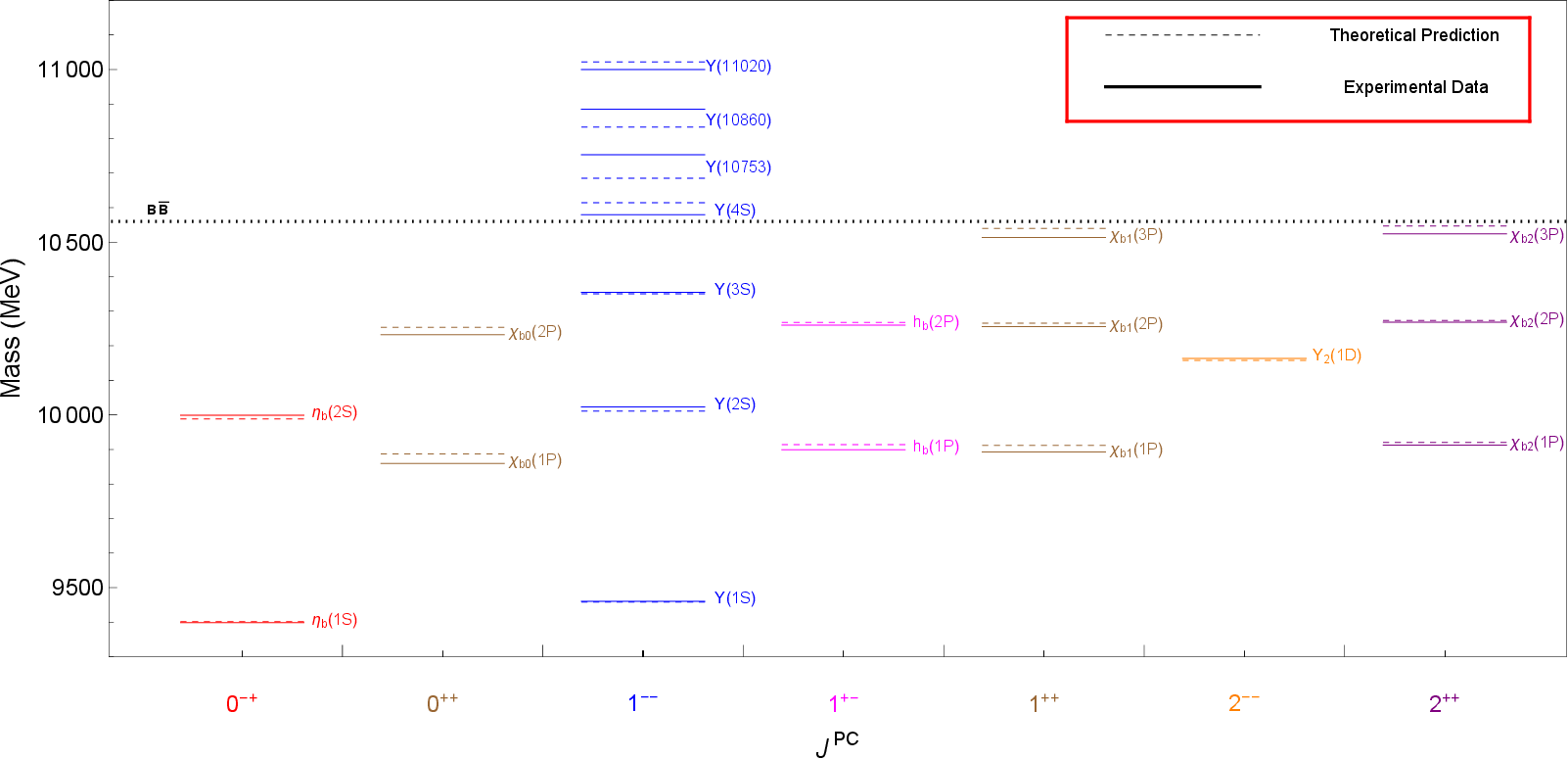} 
	\caption{Comparison of mass spectra of bottomonium obtained from our model with experimental data.}
	\label{F2}
\end{figure}

We have evaluated the mass spectra of $S$, $P$, and $D$-wave bottomonium states and have compared our results with available experimental and other theoretical models in Tables \ref{T3}, \ref{T4}, and \ref{T5} respectively. Graphical comparison of our results with PDG data \cite{23} are presented in Figure \ref{F2}. The $1S$, $2S$, $1P$, and $2P$ states are experimentally well established states and we can observe from Tables \ref{T3}, \ref{T4}, and Figure \ref{F2} that the masses predicted from our model for these states allign close to the experimental values. The experimental mass difference between $\Upsilon (1S)$ and $\eta_b (1S)$ is $62.3 \pm 3.2$ MeV \cite{23} and we obtain this difference as $57$ MeV. The Babar collaboration \cite{18} reported the mass difference between $\chi_{b1} (1P)$ and $\chi_{b0} (1P)$ as $32.5 \pm 0.9$ MeV and from our model, we obtain this difference as $25$ MeV. The reported mass difference between $\chi_{b2} (1P)$ and $\chi_{b1} (1P)$ by PDG \cite{23} is $19.10 \pm 0.25$ MeV and we obtain this difference as $8$ MeV. The mass difference between $\Upsilon(2S)$ and $\eta_b(2S)$ reported by Belle Collaboration \cite{12} which also reported the first evidence of $\eta_b(2S)$ is $24.3 \pm 3.5_{-1.9}^{+2.8}$ MeV and from our model, we obtain this difference as $22$ MeV, which is well within the experimental uncertainty. The reported mass difference between $\chi_{b1} (2P)$ and $\chi_{b0} (2P)$ by BaBar collaboration \cite{17} is $23.8 \pm 1.7$ MeV and we obtain this difference as $13$ MeV. The mass difference between $\chi_{b2} (2P)$ and $\chi_{b1} (2P)$ reported by PDG \cite{23} is $13.10 \pm 0.24$ MeV and our model gives this difference as $7$ MeV. Under $3S$ and $4S$ states, we only have spin triplet states discovered, $\Upsilon(3S)$ and $\Upsilon(4S)$ respectively with the experimental mass of $10355.1 \pm 0.5$ MeV and $10579.4 \pm 1.2$ MeV respectively \cite{23}. The predicted masses for these states from our model are $10351$ MeV and $10614$ MeV respectively. The mass difference reported by BaBar collaboration \cite{18} between $\Upsilon (3S)$ and $\Upsilon (2S)$ is $331.50 \pm 0.02 \pm 0.13$ MeV and from our model, this difference is $340$ MeV. Under $3P$ states, we only have $\chi_{b1} (3P)$ and $\chi_{b2} (3P)$ with experimental masses of $10513.4 \pm 0.7$ MeV and $10524.0 \pm 0.8$ MeV respectively \cite{23}. The predicted masses of these states from our model are $10540$ MeV and $10547$ MeV respectively. For $D$-wave bottomonium state, we only have $1$ experimentally observed state $\Upsilon_2 (1D)$ corresponding to $J^{PC}=2^{--}$ which is a $1 {}^3 D_2$ state. The experimental mass of $\Upsilon_2 (1D)$ is $10163.7 \pm 1.4$ MeV \cite{23}. The predicted mass for this state from our model is $10158$ MeV.

\begin{table}
	\caption{\label{T3} Mass Spectra of $S$-wave Bottomonium in MeV.}
	\begin{ruledtabular}
		\begin{tabular}{@{}cccccccccc}
			State&$J^{PC}$&Meson&PDG \cite{23} &Our Work&\cite{36}&\cite{35}&\cite{72}&\cite{73}&\cite{74}\\
			\hline
			$1\, {}^1S_0$&$0^{-+}$&$\eta_b (1S)$&$9398.7\pm2.0$&$9401$&$9389$&$9399$&$9423$&$9412$&$9547$\\
			$2\, {}^1S_0$&$0^{-+}$&$\eta_b (2S)$&$9999\pm4$&$9989$&$9987$&$9986$&$9983$&$9995$&$9977$\\
			$3\, {}^1S_0$&$0^{-+}$&&&$10335$&$10330$&$10315$&$10342$&$10339$&$10232$\\
			$4\, {}^1S_0$&$0^{-+}$&&&$10602$&$10595$&$10583$&$10638$&$10572$&$10436$\\
			$5\, {}^1S_0$&$0^{-+}$&&&$10823$&$10817$&$10816$&$10901$&$10746$&$10607$\\
			$6\, {}^1S_0$&$0^{-+}$&&&$11014$&$11011$&$11024$&$11140$&$11064$&\\
			$1\, {}^3S_1$&$1^{--}$&$\Upsilon(1S)$&$9460.40\pm0.10$&$9458$&$9460$&$9470$&$9463$&$9460$&$9551$\\
			$2\, {}^3S_1$&$1^{--}$&$\Upsilon(2S)$&$10023.4\pm0.5$&$10011$&$10016$&$10033$&$10001$&$10026$&$9978$\\
			$3\, {}^3S_{1}$&$1^{--}$&$\Upsilon(3S)$&$10355.1\pm0.5$&$10351$&$10351$&$10352$&$10354$&$10364$&$10232$\\
			$4\, {}^3S_1$&$1^{--}$&$\Upsilon(4S)$&$10579.4\pm1.2$&$10614$&$10611$&$10615$&$10650$&$10594$&$10437$\\
			$5\, {}^3S_1$&$1^{--}$&$\Upsilon(10860)$&$10885.2^{+2.6}_{-1.6}$&$10833$&$10831$&$10845$&$10912$&$10766$&$10607$\\
			$6\, {}^3S_1$&$1^{--}$&$\Upsilon(11020)$&$11000\pm4$&$11022$&$11023$&$11051$&$11151$&$11081$&\\
		\end{tabular}
	\end{ruledtabular}
\end{table}

\begin{table}
	\caption{\label{T4} Mass Spectra of $P$-wave Bottomonium in MeV.}
	\begin{ruledtabular}
		\begin{tabular}{@{}cccccccccc}
				State&$J^{PC}$&Meson&PDG \cite{23} &Our Work&\cite{36}&\cite{35}&\cite{72}&\cite{73}&\cite{74}\\
			\hline
			$1\, {}^1 P_1$&$1^{+-}$&$h_b (1P)$&$9899.3\pm0.8$&$9914$&$9903$&$9864$&$9899$&$9874$&$9905$\\
			$2\, {}^1 P_1$&$1^{+-}$&$h_b (2P)$&$10259.8\pm1.2$&$10268$&$10256$&$10298$&$10268$&$10270$&$10167$\\
			$3\, {}^1 P_1$&$1^{+-}$&&&$10542$&$10529$&$10555$&$10570$&$10526$&$10377$\\
			$4\, {}^1 P_1$&$1^{+-}$&&&$10769$&$10757$&$10785$&&$10714$&$10556$\\
			$5\, {}^1 P_1$&$1^{+-}$&&&$10964$&$10955$&$10994$&&$10863$&\\
			$6\, {}^1 P_1$&$1^{+-}$&&&$11136$&&&&&\\
			$1\, {}^3 P_0$&$0^{++}$&$\chi_{b0} (1P)$&$9859.44\pm0.42\pm0.31$&$9887$&$9865$&$9837$&$9874$&$9849$&$9892$\\
			$2\, {}^3 P_0$&$0^{++}$&$\chi_{b0} (2P)$&$10232.5\pm0.4\pm0.5$&$10253$&$10226$&$10258$&$10248$&$10252$&$10157$\\
			$3\, {}^3 P_0$&$0^{++}$&&&$10529$&$10502$&$10503$&$10551$&$10512$&$10368$\\
			$4\, {}^3 P_0$&$0^{++}$&&&$10757$&$10732$&$10727$&&$10703$&$10549$\\
			$5\, {}^3 P_0$&$0^{++}$&&&$10953$&$10933$&$10930$&&$10853$&\\
			$6\, {}^3 P_0$&$0^{++}$&&&$11126$&&&&&\\
			$1\, {}^3 P_1$&$1^{++}$&$\chi_{b1} (1P)$&$9892.78\pm0.26\pm0.31$&$9912$&$9897$&$9852$&$9894$&$9871$&$9904$\\
			$2\, {}^3 P_1$&$1^{++}$&$\chi_{b1} (2P)$&$10255.46\pm0.22\pm0.50$&$10266$&$10251$&$10279$&$10265$&$10267$&$10166$\\
			$3\, {}^3 P_1$&$1^{++}$&$\chi_{b1}(3P) $&$10513.4\pm0.7$&$10540$&$10524$&$10529$&$10567$&$10524$&$10375$\\
			$4\, {}^3 P_1$&$1^{++}$&&&$10767$&$10753$&$10756$&&$10713$&$10555$\\
			$5\, {}^3 P_1$&$1^{++}$&&&$10963$&$10951$&$10962$&&$10861$&\\
			$6\, {}^3 P_1$&$1^{++}$&&&$11134$&&&&&\\
			$1\, {}^3 P_2$&$2^{++}$&$\chi_{b2} (1P)$&$9912.21\pm0.26\pm0.31$&$9920$&$9918$&$9877$&$9907$&$9881$&$9911$\\
			$2\, {}^3 P_2$&$2^{++}$&$\chi_{b2} (2P)$&$10268.65\pm0.22\pm0.50$&$10273$&$10269$&$10317$&$10274$&$10274$&$10171$\\
			$3\, {}^3 P_2$&$2^{++}$&$\chi_{b2} (3P)$&$10524.0\pm0.8$&$10547$&$10540$&$10580$&$10576$&$10530$&$10380$\\
			$4\, {}^3 P_2$&$2^{++}$&&&$10774$&$10767$&$10814$&&$10717$&$10559$\\
			$5\, {}^3 P_2$&$2^{++}$&&&$10969$&$10965$&$11026$&&$10865$&\\
			$6\, {}^3 P_2$&$2^{++}$&&&$11140$&&&&&\\
		\end{tabular}
	\end{ruledtabular}
\end{table}

\begin{table}
	\caption{\label{T5} Mass Spectra of $D$-wave Bottomonium in MeV.}
	\begin{ruledtabular}
		\begin{tabular}{@{}cccccccccc}
			State&$J^{PC}$&Meson&PDG \cite{23} &Our Work&\cite{36}&\cite{35}&\cite{72}&\cite{73}&\cite{74}\\
			\hline
			$1\, {}^1D_2$&$2^{-+}$&&&$10157$&$10152$&$10140$&$10149$&$10153$&$10089$\\
			$2\, {}^1D_2$&$2^{-+}$&&&$10448$&$10439$&$10519$&$10465$&$10456$&$10306$\\
			$3\, {}^1D_2$&$2^{-+}$&&&$10687$&$10677$&$10733$&$10740$&$10664$&$10493$\\
			$4\, {}^1D_2$&$2^{-+}$&&&$10892$&$10883$&$10940$&$10988$&$10823$&$10650$\\
			$5\, {}^1D_2$&$2^{-+}$&&&$11071$&$11066$&$11145$&&$10952$&\\
			$6\, {}^1D_2$&$2^{-+}$&&&$11229$&&&&$11057$&\\
			$1\, {}^3D_1$&$1^{--}$&&&$10156$&$10145$&$10086$&$10145$&$10144$&$10086$\\
			$2\, {}^3D_1$&$1^{--}$&&&$10446$&$10432$&$10451$&$10462$&$10450$&$10303$\\
			$3\, {}^3D_1$&$1^{--}$&$\Upsilon(10753)$&$10753\pm6$&$10685$&$10670$&$10652$&$10736$&$10659$&$10490$\\
			$4\, {}^3D_1$&$1^{--}$&&&$10890$&$10877$&$10848$&$10985$&$10818$&$10648$\\
			$5\, {}^3D_1$&$1^{--}$&&&$11069$&$11060$&$11047$&&$10949$&\\
			$6\, {}^3D_1$&$1^{--}$&&&$11228$&&&&$11054$&\\
			$1\, {}^3D_2$&$2^{--}$&$\Upsilon_2(1D)$&$10163.7\pm1.4$&$10158$&$10151$&$10123$&$10149$&$10152$&$10089$\\
			$2\, {}^3D_2$&$2^{--}$&&&$10448$&$10438$&$10497$&$10465$&$10455$&$10306$\\
			$3\, {}^3D_2$&$2^{--}$&&&$10687$&$10676$&$10707$&$10740$&$10664$&$10493$\\
			$4\, {}^3D_2$&$2^{--}$&&&$10892$&$10882$&$10909$&$10988$&$10822$&$10650$\\
			$5\, {}^3D_2$&$2^{--}$&&&$11071$&$11065$&$11113$&&$10951$&\\
			$6\, {}^3D_2$&$2^{--}$&&&$11229$&&&&$11057$&\\
			$1\, {}^3D_3$&$3^{--}$&&&$10157$&$10156$&$10175$&$10150$&$10158$&$10091$\\
			$2\, {}^3D_3$&$3^{--}$&&&$10449$&$10442$&$10563$&$10466$&$10459$&$10308$\\
			$3\, {}^3D_3$&$3^{--}$&&&$10688$&$10680$&$10787$&$10741$&$10667$&$10495$\\
			$4\, {}^3D_3$&$3^{--}$&&&$10893$&$10886$&$11000$&$10990$&$10825$&$10651$\\
			$5\, {}^3D_3$&$3^{--}$&&&$11072$&$11069$&$11211$&&$10954$&\\
			$6\, {}^3D_3$&$3^{--}$&&&$11230$&&&&$11059$&\\
		\end{tabular}
	\end{ruledtabular}
\end{table}

\begin{table}
	\caption{Mass difference between $nD$ state and $(n+1)S$ state in MeV corresponding to $J^{PC}=1^{--}$ \label{T6}}
	\begin{ruledtabular}
		\begin{tabular}{@{}cc}
			$nD-(n+1)S$&Mass Difference\\
			\hline
			$1D-2S$&$145$\\
			$2D-3S$&$95$\\
			$3D-4S$&$71$\\
			$4D-5S$&$57$\\
			$5D-6S$&$47$\\
		\end{tabular}
	\end{ruledtabular}
\end{table}

We have assigned $3$ experimentally observed states by comparing their $J^{PC}$ values and their observed masses with those predicted from our model. $\Upsilon(10753)$ with observed mass of $10753 \pm 6$ MeV \cite{23} is assigned as $3 {}^3D_1$ state with predicted mass of $10685$ MeV. We have assigned $\Upsilon(10860)$ with observed mass of $10885.2_{-1.6}^{+2.6}$ MeV \cite{23} as $5 {}^3S_1$ state with predicted mass of $10833$ MeV. $\Upsilon(11020)$ with experimental mass of $11000 \pm 4$ MeV \cite{23} is assigned as $6 {}^3S_1$ state with predicted mass of $11022$ MeV. We can observe from Figure \ref{F2} that below $B\overline{B}$ threshold, the predicted masses from our model lies close to the experimental values, whereas above this threshold, there are slight deviations. Above $B\overline{B}$ threshold, there are at present $4$ experimental states $\Upsilon(4S)$, $\Upsilon(10860)$, $\Upsilon(11020)$, and $\Upsilon(10753)$. All these states corresponds to $J^{PC}=1^{--}$ with the radial quantum number $n\geq3$. These deviations arises because of the fact that there can be mixing between $(n+1)S$ state and $nD$ state ($S-D$ mixing) \cite{69,70,91} which we have not considered in this work. This mixing becomes significant for $n\geq3$ \cite{69} and hence we can observe that only for $n\geq3$, the deviations are large whereas for $n<3$, the deviations are negligible for $J^{PC}=1^{--}$. There can also be coupled channel effects \cite{71} above $B\overline{B}$ threshold, that can be responsible for these observed deviations. It is found in Ref. \cite{69} that the difference in mass between $nD$ state and $(n+1)S$ state is small and that the mass difference decreases with increase in $n$. This can be found in Table \ref{T6}, where we have tabulated the mass difference between $nD$ state and $(n+1)S$ state corresponding to $J^{PC}=1^{--}$ obtained from our model.

\subsection{$bb$ diquarks:}
\begin{figure}
	\centering
	\includegraphics[width=1\linewidth]{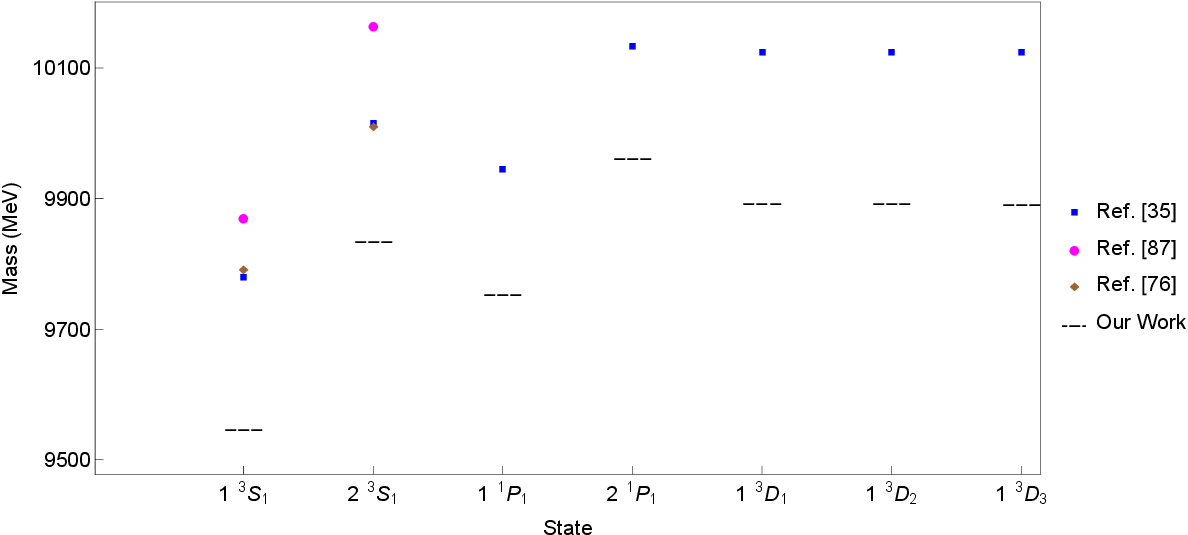} 
	\caption{Comparison of mass spectra of $bb$-diquark obtained from our model with other theoretical models in MeV.}
	\label{F1}
\end{figure}

We have computed the mass spectra of $S$, $P$, and $D$-wave $bb$ diquarks and have compared with other theoretical models and with the corresponding bottomonium states evaluated from our model in Table \ref{T7}. In Figure \ref{F1}, we have shown the comparison of the masses of $bb$-diquarks obtained from our model with other theoretical models. We can observe from Table \ref{T7} that the masses of diquark obtained from our model is around $200$ MeV less than those obtained from other models. For the excited states, the reason for this difference could be the use of different confinement potentials. However, in the case of lower states, it is evident that the Coulomb interaction dominates across all the models and hence the difference should be minimal. Nonetheless, in our model, we have incorporated $O(1/m)$ interactions which also significantly influence the lower states. This could be the reason for the discrepancy observed in the lower states. If we were to compare the diquark masses with corresponding bottomonium states, we can observe that except for $1 {}^3S_1$ state, the $bb$-diquark mass is significantly less than the corresponding bottomonium states. This is because of the fact that for $1 {}^3S_1$ state, the interaction is dominated by Coulomb interaction which is $-\frac{2 \alpha_s}{3r}$ for $bb$-diquark and $-\frac{4\alpha_s}{3r}$ for bottomonium. This Coulomb interaction is greater for diquark resulting in higher interaction energy, thus resulting in higher mass. For excited states, the interaction is dominated by confinement part which is $\frac{\lambda (1-e^{-\nu r})}{2r}$ for $bb$-diquark and  $\frac{\lambda (1-e^{-\nu r})}{r}$ for bottomonium. Here, the interaction energy will be lower for diquark resulting in lesser mass compared to corresponding bottomonium states.

\begin{table}
	\caption{\label{T7}Mass Spectra of $bb$ diquark in MeV.}
	\begin{ruledtabular}
		\begin{tabular}{@{}cccccc}
			State&Our Work&\cite{33}&\cite{77}&\cite{65}&$b\overline{b}$ (our work)\\
			\hline
			$1\, {}^3S_1$&$9548$&$9778$&$9871$&$9792$&$9458$\\
			$2\, {}^3S_1$&$9836$&$10015$&$10165$&$10011$&$10011$\\
			$1\, {}^1P_1$&$9755$&$9944$&&&$9914$\\
			$2\, {}^1P_1$&$9964$&$10132$&&&$10268$\\
			$1\, {}^3D_1$&$9894$&$10123$&&&$10156$\\
			$2\, {}^3D_1$&$10066$&&&&$10446$\\
			$1\, {}^3D_2$&$9894$&$10123$&&&$10158$\\
			$2\, {}^3D_2$&$10066$&&&&$10448$\\
			$1\, {}^3D_3$&$9893$&$10123$&&&$10157$\\
			$2\, {}^3D_3$&$10066$&&&&$10448$\\
		\end{tabular}
	\end{ruledtabular}
\end{table}

\subsection{Triply bottom baryons:}
\begin{figure}
	\centering
	\includegraphics[width=0.9\linewidth]{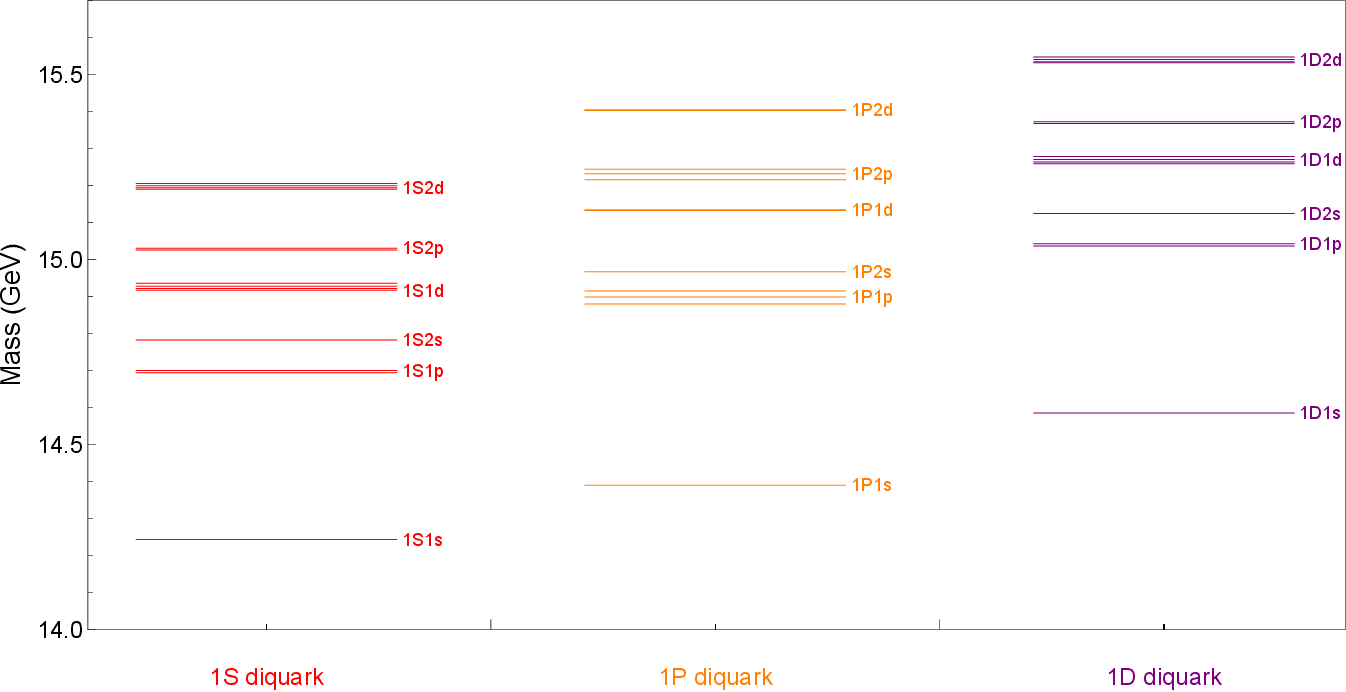} 
	\caption{Mass spectra of triply bottom baryons corresponding to different diquark states with $n_d=1$.}
	\label{F4}
\end{figure}

\begin{figure}
	\centering
	\includegraphics[width=0.9\linewidth]{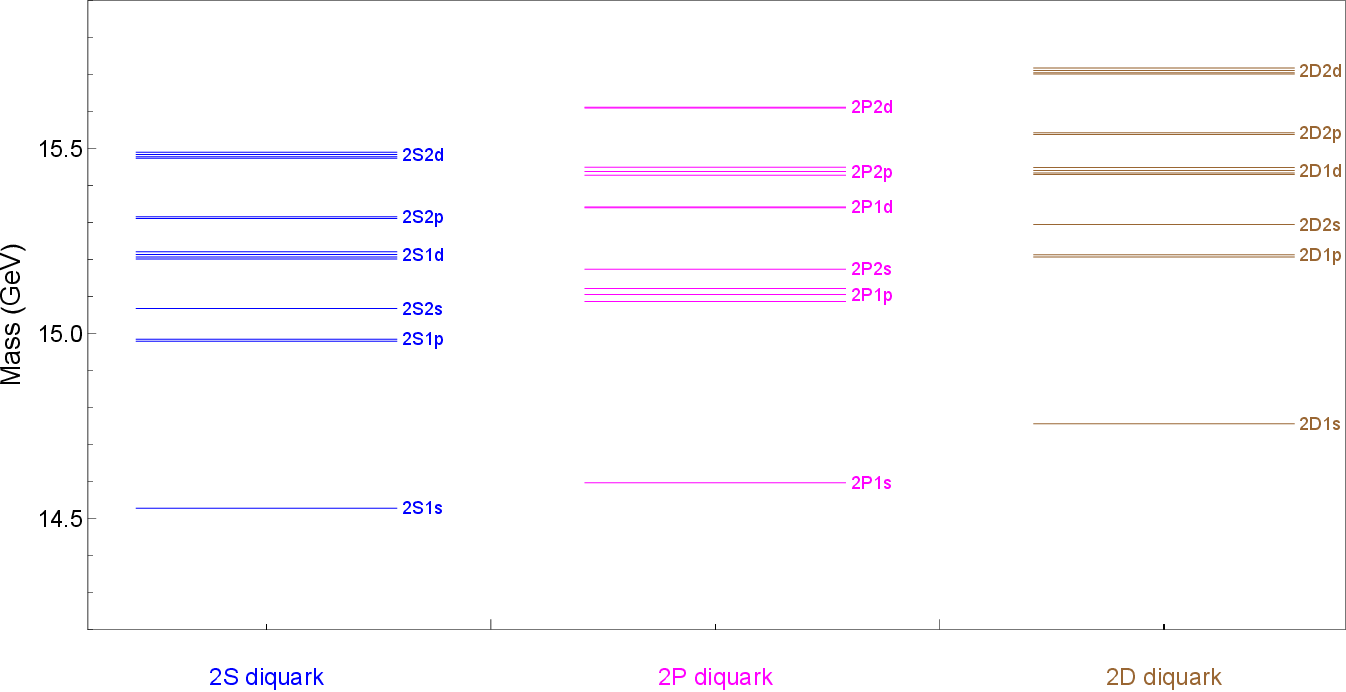} 
	\caption{Mass spectra of triply bottom baryons corresponding to different diquark states with $n_d=2$.}
	\label{F5}
\end{figure}

\begin{figure}
	\centering
	\includegraphics[width=0.9\linewidth]{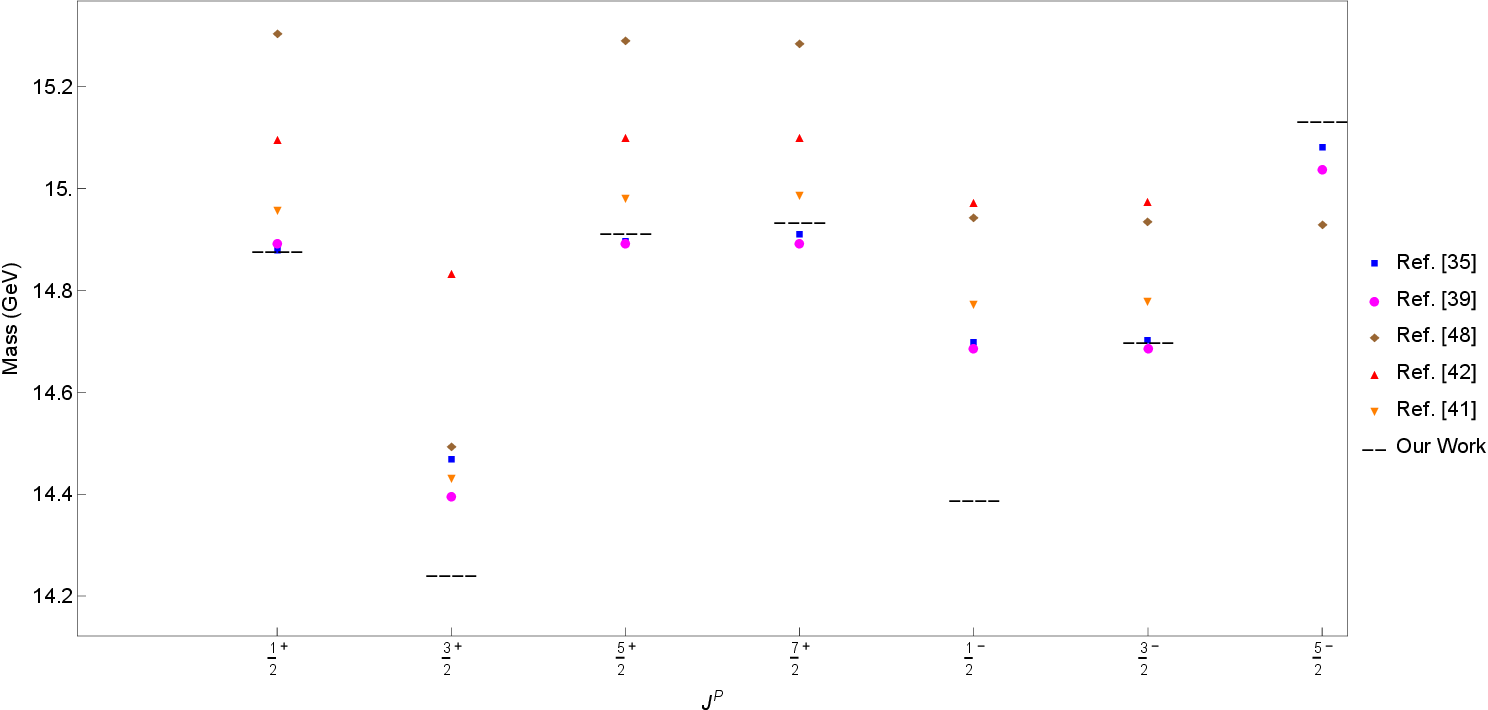} 
	\caption{Comparison of masses of lowest state triply bottom baryons corresponding to different $J^P$ values in GeV.}
	\label{F6}
\end{figure}

We have computed the mass spectra of triply bottom baryons using the diquark-quark model. The triply bottom baryons are modelled as a bound state of a $bb$-diquark and a $b$-quark. We have considered the radial and orbital excitations of both diquark and diquark-quark system. The computed mass spectra are presented in Tables \ref{T8} and \ref{T9}. In Figures \ref{F4} and \ref{F5}, we have presented the mass spectra of triply bottom baryons corresponding to different diquark states with $n_d=1$ and $n_d=2$ respectively. In Table \ref{T10}, we have compared the masses of ground state triply bottom baryons obtained from our model with other theoretical models. We can observe that the ground state mass varies in between $13.200$ GeV to $15.200$ GeV across different models and the ground state mass obtained from our model is $14.243$ GeV. In Table \ref{T11} and Figure \ref{F6}, we have compared the masses of lowest state triply bottom baryons corresponding to $J^P=\frac{1}{2}^+, \frac{3}{2}^+, \frac{5}{2}^+, \frac{7}{2}^+, \frac{1}{2}^-, \frac{3}{2}^-,$ and $\frac{5}{2}^-$ obtained from our model with other theoretical models. The authors of Ref. \cite{33} have used diquark-quark model with linear confinement potential, Ref. \cite{37} have used constituent quark model with screened confinement potential, Ref. \cite{43} have used hypercentral constituent quark model with linear confinement potential, and Refs \cite{79} and \cite{80} have used constituent quark model with linear confinement potential. According to Ref. \cite{88}, in the doubly heavy baryons, when the diquark is formed by the combination of two heavy quarks, for a given $\mathbf{J_d}$ and $\mathbf{j_q}$, the masses are degenerate for different $\mathbf{J}$ values of baryon. We can observe from Tables \ref{T8} and \ref{T9} and from Figures \ref{F4} and \ref{F5} that even in triply heavy baryons, for a given $\mathbf{J_d}$ and $\mathbf{j_q}$, the masses are almost degenerate (negligible difference in mass) for different $\mathbf{J}$ values of baryon.

\begin{table}
	\caption{\label{T8}Mass Spectra of $bbb$ baryon in GeV (positive parity). }
	\begin{ruledtabular}
		\begin{tabular}{@{}cccccccccccc}
			State&$J^P$&Mass&State&$J^P$&Mass&State&$J^P$&Mass&State&$J^P$&Mass\\
			\hline
			$1S1s$&$\frac{3}{2}^+$&$14.243$&$1S2s$&$\frac{3}{2}^+$&$14.783$&$2S1s$&$\frac{3}{2}^+$&$14.528$&$2S2s$&$\frac{3}{2}^+$&$15.068$\\
			$1S1d$&$\frac{1}{2}^+$&$14.917$&$1S2d$&$\frac{1}{2}^+$&$15.190$&$2S1d$&$\frac{1}{2}^+$&$15.202$&$2S2d$&$\frac{1}{2}^+$&$15.474$\\
			&$\frac{3}{2}^+$&$14.922$&&$\frac{3}{2}^+$&$15.194$&&$\frac{3}{2}^+$&$15.207$&&$\frac{3}{2}^+$&$15.478$\\
			&$\frac{5}{2}^+$&$14.928$&&$\frac{5}{2}^+$&$15.200$&&$\frac{5}{2}^+$&$15.214$&&$\frac{5}{2}^+$&$15.484$\\
			&$\frac{7}{2}^+$&$14.936$&&$\frac{7}{2}^+$&$15.206$&&$\frac{7}{2}^+$&$15.221$&&$\frac{7}{2}^+$&$15.490$\\
			$1P1p$&$\frac{1}{2}^+$&$14.880$&$1P2p$&$\frac{1}{2}^+$&$15.216$&$2P1p$&$\frac{1}{2}^+$&$15.087$&$2P2p$&$\frac{1}{2}^+$&$15.428$\\
			&$\frac{3}{2}^+$&$14.899$&&$\frac{3}{2}^+$&$15.232$&&$\frac{3}{2}^+$&$15.106$&&$\frac{3}{2}^+$&$15.438$\\
			&$\frac{5}{2}^+$&$14.915$&&$\frac{5}{2}^+$&$15.244$&&$\frac{5}{2}^+$&$15.122$&&$\frac{5}{2}^+$&$15.450$\\
			$1D1s$&$\frac{3}{2}^+$&$14.585$&$1D2s$&$\frac{3}{2}^+$&$15.125$&$2D1s$&$\frac{3}{2}^+$&$14.756$&$2D2s$&$\frac{3}{2}^+$&$15.295$\\
			$1D1d$&$\frac{1}{2}^+$&$15.259$&$1D2d$&$\frac{1}{2}^+$&$15.532$&$2D1d$&$\frac{1}{2}^+$&$15.430$&$2D2d$&$\frac{1}{2}^+$&$15.702$\\
			&$\frac{3}{2}^+$&$15.264$&&$\frac{3}{2}^+$&$15.535$&&$\frac{3}{2}^+$&$15.434$&&$\frac{3}{2}^+$&$15.705$\\
			&$\frac{5}{2}^+$&$15.271$&&$\frac{5}{2}^+$&$15.541$&&$\frac{5}{2}^+$&$15.441$&&$\frac{5}{2}^+$&$15.711$\\
			&$\frac{7}{2}^+$&$15.279$&&$\frac{7}{2}^+$&$15.548$&&$\frac{7}{2}^+$&$15.449$&&$\frac{7}{2}^+$&$15.718$\\
		\end{tabular}
	\end{ruledtabular}
\end{table}

\begin{table}
	\caption{\label{T9}Mass Spectra of $bbb$ baryon in GeV (negative parity).}
	\begin{ruledtabular}
		\begin{tabular}{@{}cccccccccccc}
			State&$J^P$&Mass&State&$J^P$&Mass&State&$J^P$&Mass&State&$J^P$&Mass\\
			\hline
			$1S1p$&$\frac{1}{2}^-$&$14.694$&$1S2p$&$\frac{1}{2}^-$&$15.026$&$2S1p$&$\frac{1}{2}^-$&$14.979$&$2S2p$&$\frac{1}{2}^-$&$15.311$\\
			&$\frac{3}{2}^-$&$14.700$&&$\frac{3}{2}^-$&$15.031$&&$\frac{3}{2}^-$&$14.985$&&$\frac{3}{2}^-$&$15.316$\\
			$1P1s$&$\frac{1}{2}^-$&$14.390$&$1P2s$&$\frac{1}{2}^-$&$14.967$&$2P1s$&$\frac{1}{2}^-$&$14.597$&$2P2s$&$\frac{1}{2}^-$&$15.174$\\
			$1P1d$&$\frac{3}{2}^-$&$15.133$&$1P2d$&$\frac{3}{2}^-$&$15.403$&$2P1d$&$\frac{3}{2}^-$&$15.340$&$2P2d$&$\frac{3}{2}^-$&$15.610$\\
			&$\frac{5}{2}^-$&$15.135$&&$\frac{5}{2}^-$&$15.405$&&$\frac{5}{2}^-$&$15.342$&&$\frac{5}{2}^-$&$15.612$\\
			$1D1p$&$\frac{1}{2}^-$&$15.037$&$1D2p$&$\frac{1}{2}^-$&$15.368$&$2D1p$&$\frac{1}{2}^-$&$15.207$&$2D2p$&$\frac{1}{2}^-$&$15.538$\\
			&$\frac{3}{2}^-$&$15.043$&&$\frac{3}{2}^-$&$15.373$&&$\frac{3}{2}^-$&$15.213$&&$\frac{3}{2}^-$&$15.543$\\
		\end{tabular}
	\end{ruledtabular}
\end{table}

For $J^P=\frac{1}{2}^+$, the lowest mass obtained from our model is $14.880$ GeV ($1P1p$ state) which is close to the mass obtained by Ref.s \cite{33, 17, 79} but less than around $426$ MeV and $217$ MeV from the mass obtained in Ref. \cite{43} and \cite{80} respectively. The lowest state mass corresponding to $J^P=\frac{3}{2}^+$ corresponds to ground state mass of triply bottom baryons and our estimates are compared with other theoretical models in Table \ref{T10}. The lowest state mass corresponding to $J^P=\frac{5}{2}^+$ predicted from our model is $14.915$ GeV ($1P1p$ state) which is close to the mass predicted in Refs \cite{33, 17, 79} but is less than around $378$ MeV and $186$ MeV from the mass predicted in Ref. \cite{43} and \cite{80} respectively. The lowest state mass corresponding to $J^P=\frac{7}{2}^+$ obtained from our model is $14.936$ GeV ($1S1d$ state) which is close to the mass obtained in Refs \cite{33, 17, 79} but is less than around $350$ MeV and $165$ MeV from the mass obtained in Ref. \cite{43} and \cite{80} respectively. These differences in mass predictions may arise due to the differences in approaches employed and the differences in model parameters. Both in Refs \cite{79} and \cite{80}, the masses of triply bottom baryons have been estimated within the constituent quark model with a linear confinement potential, but due to the differences in the parameters used, there are variations between the masses predicted from these models. The model parameters in Ref. \cite{79} are obtained by fitting into the experimentally observed bottomonium states whereas in Ref. \cite{80}, the model parameters are obtained by fitting the experimentally observed singly bottom baryons. For $J^P=\frac{1}{2}^-$, the lowest state mass obtained from our model is $14.390$ GeV ($1P1s$ state) which is less than those predicted from other models. For $J^P=\frac{3}{2}^-$, the lowest state mass obtained from our model is $14.700$ GeV ($1S1p$ state) which is close to the mass obtained in Refs \cite{33, 17, 79}. For $J^P=\frac{5}{2}^-$, the lowest mass obtained from our model is $15.135$ GeV ($1P1d$ state) which is close to the mass predicted by Ref.s \cite{33, 17, 79}. However, the validity of these theoretical models hinges on the experimental discovery of these states.

\begin{table}
	\caption{\label{T10}Comparison of ground state mass of triply bottom baryons with other theoretical approaches in GeV.}
	\begin{ruledtabular}
		\begin{tabular}{@{}cc}
			Approach&Mass\\
			\hline
			Constituent Quark Model \cite{37}&$14.396$\\
			Constituent Quark Model \cite{79}&$14.432$\\
			Constituent Quark Model \cite{80}&$14.834$\\
			Effective Quark Model \cite{85}&$15.129$\\
			Relativistic Quark Model \cite{78}&$14.569$\\
			Hypercentral Constituent Quark Model \cite{43}&$14.496$\\
			Faddev Equation \cite{83}&$14.370$\\
			Regge Phenomenology \cite{84}&$14.822$\\
			QCD Sum Rules \cite{39}&$13.28 \pm 0.10$\\
			QCD Sum Rules \cite{40}&$14.3 \pm 0.2$\\
			QCD Sum Rules \cite{81}&$14.43 \pm 0.009$\\
			Diquark-quark Model \cite{33}&$14.468$\\
			Diquark-quark Model \cite{86}&$14.370$\\
			\textbf{Our Work}&$\mathbf{14.243}$\\
		\end{tabular}
	\end{ruledtabular}
\end{table}

\begin{table}
	\caption{\label{T11}Comparison of masses of lowest state triply bottom baryons corresponding to different $J^P$ values in GeV.}
	\begin{ruledtabular}
		\begin{tabular}{@{}ccccccc}
			$J^P$&Our Work&\cite{33}&\cite{37}&\cite{43}&\cite{80}&\cite{79}\\
			\hline
			$\frac{1}{2}^{+}$&$14.880$&$14.877$&$14.894$&$15.306$&$15.097$&$14.959$\\
			$\frac{3}{2}^{+}$&$14.243$&$14.468$&$14.396$&$14.496$&$14.834$&$14.432$\\
			$\frac{5}{2}^{+}$&$14.915$&$14.895$&$14.894$&$15.293$&$15.101$&$14.981$\\
			$\frac{7}{2}^{+}$&$14.936$&$14.909$&$14.894$&$15.286$&$15.101$&$14.988$\\
			$\frac{1}{2}^{-}$&$14.390$&$14.698$&$14.688$&$14.944$&$14.975$&$14.773$\\
			$\frac{3}{2}^{-}$&$14.700$&$14.702$&$14.688$&$14.937$&$14.976$&$14.779$\\
			$\frac{5}{2}^{-}$&$15.135$&$15.081$&$15.038$&$14.931$&&\\
		\end{tabular}
	\end{ruledtabular}
\end{table}

\section{Summary: \label{S4}}
In this work, we have evaluated the mass spectrum of bottomonium, $bb$-diquark, and triply bottom baryons by utilizing a phenomenological potential model comprising of short-range one-gluon Coulomb potential, a screened confinement potential, and $O(1/m)$ corrections predicted from pNRQCD and lattice studies. Unlike most models that use linear confinement potential, we have applied a screened confinement potential along with $O(1/m)$ correction terms to predict the mass spectra of beauty hadrons. This approach takes into account several features, including the flattening of the linear confinement potential as the separation between quarks increases, as well as the effects of finite heavy quark masses. The model parameters are obtained by fitting some of the experimentally well-established bottomonium states. The predicted masses of bottomonium are compared with other theoretical models and available experimental results. Below the $B\overline{B}$ threshold, the mass predicted from our model is close to the experimental values, whereas above this threshold, there are small deviations from the experimental results. The mass spectra of triply bottom baryons are evaluated within the diquark-quark model, and we have compared them with other theoretical models, as there are no experimental states discovered yet. Our predictions can help to narrow down the expected mass range for the experimental discovery of triply bottom baryons. 

\begin{acknowledgements}
	One of the authors (RK) is grateful to the Manipal Academy of Higher Education, Manipal for financial support under ‘Dr. TMA Pai Scholarship Programme’.
\end{acknowledgements}

\bibliographystyle{apsrev4-2}
\bibliography{bbb}

\end{document}